\documentclass[10pt,journal,compsoc]{IEEEtran}
\ifCLASSOPTIONcompsoc
  \usepackage[nocompress]{cite}
\else
  \usepackage{cite}
\fi
\ifCLASSINFOpdf
\else
\fi

\usepackage[english]{babel} 

\usepackage{graphicx} 
\usepackage{epstopdf}

\usepackage{subfigure}

\usepackage{wrapfig} 

\usepackage{booktabs} 

\usepackage{algorithm}  
\usepackage{algpseudocode}  
\usepackage{amsmath}  

\usepackage{enumitem} 
\setlist{noitemsep} 

\makeatletter
\renewcommand\@biblabel[1]{\textbf{#1.}} 
\makeatother

\usepackage{mathrsfs}
\usepackage{amsmath}
\usepackage{amsfonts}
\usepackage{amsthm}

\begin{document}
%
\title{Representation Learning of Graphs \\ Using Graph Convolutional Multilayer \\ Networks Based on Motifs}
%
%
%
%

\author{Xing~Li,
        Wei~Wei,
        Xiangnan~Feng,
		Xue~Liu,
		Zhiming~Zheng
\IEEEcompsocitemizethanks{\IEEEcompsocthanksitem W. Wei was with the School of Mathematical Science, Beihang University, Beijing, China, Key Laboratory of Mathematics Informatics Behavioral Semantics, Ministry of Education, China, Peng Cheng Laboratory, Shenzhen, Guangdong, China and Beijing Advanced Innovation Center for Big Data and Brain Computing, Beihang University, Beijing, China. \protect\\
E-mail: see weiw@buaa.edu.cn
\IEEEcompsocthanksitem X. Li, X. Feng, X. Liu and Z. Zheng were with the School of Mathematical Science, Beihang University, Beijing, China, Key Laboratory of Mathematics Informatics Behavioral Semantics, Ministry of Education, China and Beijing Advanced Innovation Center for Big Data and Brain Computing, Beihang University, Beijing, China.}
\thanks{ }}

\IEEEtitleabstractindextext{%
\begin{abstract}
The graph structure is a commonly used data storage mode, and it turns out that the low-dimensional embedded representation of nodes in the graph is extremely useful in various typical tasks, such as node classification, link prediction , etc. However, most of the existing approaches start from the binary relationship (i.e., edges) in the graph and have not leveraged the higher order local structure (i.e., motifs) of the graph. Here, we propose mGCMN -- a novel framework which utilizes node feature information and the higher order local structure of the graph to effectively generate node embeddings for previously unseen data. Through research we have found that different types of networks have different key motifs. And the advantages of our method over the baseline methods have been demonstrated in a large number of experiments on citation network and social network datasets. At the same time, a positive correlation between increase of the classification accuracy and the clustering coefficient is revealed. It is believed that using high order structural information can truly manifest the potential of the network, which will greatly improve the learning efficiency of the graph neural network and promote a brand-new learning mode establishment.
\end{abstract}

}

\maketitle

\IEEEdisplaynontitleabstractindextext

%
\IEEEpeerreviewmaketitle

\IEEEraisesectionheading{\section{Introduction}\label{sec:introduction}}

%
%
%
%
\IEEEPARstart{G}{raph} structure is a common and flexible data structure that can represent data in a variety of fields, including social networks$^{1}$, biological protein-protein networks$^{2}$, knowledge network$^{3}$, etc. People can use graphs to efficiently store and access relational knowledge about interactive entities. For example, in a social network, a node can represent a person, and nodes are connected with edges indicating that people know each other. In recent years, due to the rapidly increase in the amount of data, the formed graphs have become increasingly complicated, which makes it difficult to extract valid information from complicated organizations. Therefore, it is important to process complex raw data in advance and convert them into a form that can be effectively developed for gaining good results. 

High-dimensional complex data are expected to be represented as a simple, easy-to-process low-dimensional representation. However, traditional manual feature extraction requires a great deal of manpower and relies on highly specialized knowledge. Thus, representation learning has played a key role in graph machine learning. Representation learning is a technical method to learn the characteristics of data, transforming raw data into a form that can be effectively developed using machine learning. It avoids the trouble of manually extracting features and allows the computer to learn how to extract features while learning to use features, namely, learning how to learn. Representation learning can be regarded as a type of preprocessing, does not directly obtain the results but an effective representation for producing desirable results. In other words, the choice of representation usually depends on subsequent learning tasks, i.e., a good representation should make learning of downstream tasks easier.

The main topic of representation learning on graph is to deal with the relational mode or connection pattern. For this learning, effective encoding of the basic structures such as nodes, edges, subgraphs will lead to quantitative understanding of the data knowledge, and help promote the learning efficiency combined with the downstream tasks. Recently, many valuable results have been obtained. Laplacian feature map is one of the earliest and most famous representation learning methods whose loss function weights pairs of nodes according to their proximity in the graph $^{4}$, which is a direct encoding method. DeepWalk$^{5}$ and node2vec$^{6}$ also rely on direct encoding. However, instead of attempting to decode fixed deterministic distance metrics, these methods gain the representation of the target objects through the random walk on the graph, which makes the graph proximity measure more flexible and has led to superior performance in a number of settings. However, these direct encoding methods have disadvantages such as too many parameters and insufficient use of information in the graph (such as node characteristics). Therefore, the graph neural network (GNN) framework which obtains the representation of the target object through deep learning is developed$^{7}$. Inspired by the parameter sharing operation in the convolutional neural network, the graph convolution network (GCN) is developed (Kipf et al.$^{8}$), so that the convolution operation can be applied to the irregular graph data (relative to the regular image data). However, all the above methods start from the binary relationship (i.e., edge) in the graph, and can not leverage the higher-order local structure (i.e., motifs) in the graph, which may help to explore more effective information in more complex graph structures.

\noindent\textbf{Present work.} This paper develops a new framework that combines motif with traditional representation learning. We first analyze the important statistics in the graph; then the graph convolutional neural network is chosen as the basic model and a new framework is developed. This new framework is named graph convolutional multilayer networks based on motifs (mGCMN), which can improve the accuracy of the task while spending a little more time. It is believed that combining motifs is in essence to redefine the node neighbors and redistribute the weight of the graph network. And we apply a variety of motifs and conduct a large number of experiments, all of which obtain better test results than the baselines. At the same time, the relationship between classification accuracy and clustering coefficient is revealed.

The rest of this article is organized as follows. In Section 2, related past work is outlined and in Section 3, the MGCMN which is our representation learning method, is introduced. Our experiments will be introduced in Section 4 and the results are given in Section 5. In Section 6, the relevant work and conclusions are discussed.

\begin{figure}
\center
\includegraphics[width=\linewidth]{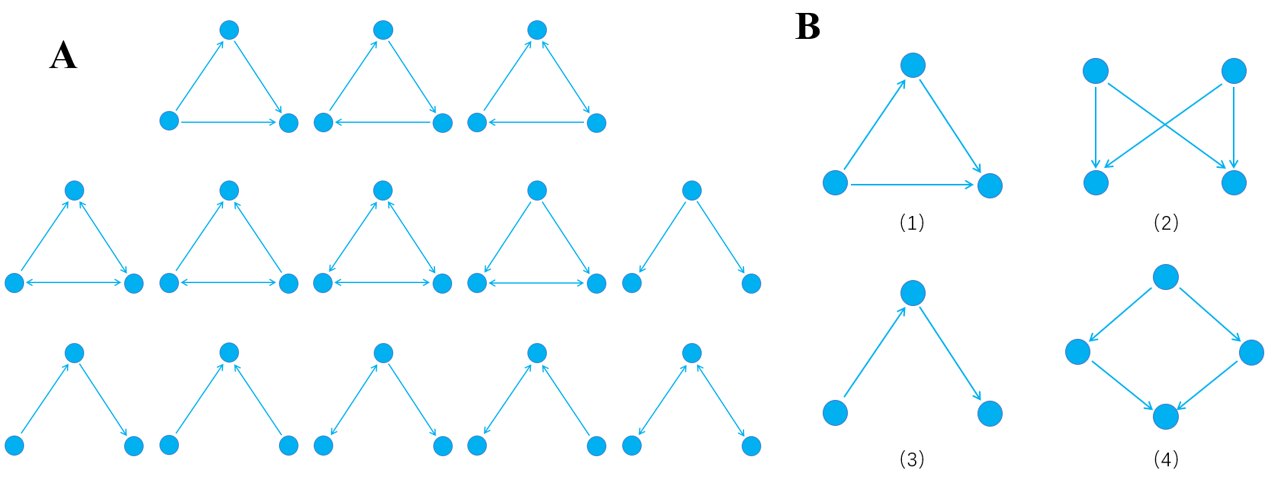}
\caption{A.Examples of 3-order network structures (3-motifs). B.The four motifs.}
\vspace{1ex}
(1) feedforward loop.  (2) bi-fan.  (3) three chain.  (4) bi-parallel.
\end{figure}

\section{Related Work}
Our method is related to recent advances in the concepts and applications of motif, as well as previous representation learning methods such as semi-supervised learning methods that apply convolutional neural networks to graph structure data. So this section will focus on the previous work closely related to mGCMN.

\subsection{The Concept and Application of Motif}
Motif is the interconnection pattern that occurs in complex networks, whose number is significantly higher than that in random networks under given conditions. It is generally considered to be the basic building blocks of a complex network. For example, Fig.1A shows all the 3-order directed motifs.

Motif is important and previous research has established that it provides a new perspective on identifying graph types. For example, two transcriptional regulatory networks (transcriptional regulatory networks are biochemical networks responsible for regulating gene expression in cells) correspond to organisms from different fields: eukaryotes (Saccharomyces cerevisiae) and bacteria (E. coli)$^{9}$. Two transcription networks show the same motifs: a three-node pattern which is called "feedforward loop" and a four-node pattern which is called "bi-fan", and general trends in the food web are shown as: a three-node pattern which is called "three chain" and a four-node pattern which is called "bi-parallel" (The four motifs are shown in Fig.1B)$^{9}$. The food web responds to energy flow, while the gene regulatory network responds to information flow, which seems to have a significantly different structure from energy flow; on the other hand, we can capture important structural information (such as geographic location information, urban hubs, etc.) that is difficult to capture through the edges by selecting the appropriate motifs$^{10}$.

\subsection{Representation Learning Method}
Representation learning is a technical method to learn the characteristics of data. It converts the original data into a low-dimensional information-rich form that is convenient for machine learning to develop effectively. From a certain perspective, it can be regarded as a dimensionality reduction method. Due to the flexibility of the graph structure, a lot of raw data is stored in the form of graphs, so the representation learning methods introduced below are all graph representation learning methods. Generally speaking, it can be divided into three categories$^{11}$:

\subsubsection{Embedding Approaches Based on Factorization}
Inspired by classic techniques for dimensionality reduction, early methods for learning representations of nodes largely focused on matrix-factorization approaches. A representative example here is: Laplacian eigenmaps method, which we can view within the encoder-decoder framework as a direct encoding approach.$^{4}$ Following the Laplacian eigenmaps method, there are a large number of representation learning methods based on inner product, such as Graph Factorization (GF)$^{12}$, GraRep$^{13}$ and HOPE$^{14}$. And the main difference of them is that the basic matrix used is different. In GF method, the original adjacency matrix of graph is used. And GraRep is based on various powers of the original adjacency matrix. As for HOPE, more general variants of the original adjacency matrix are considered.

\begin{figure*}
\begin{center}
\includegraphics[width=\linewidth]{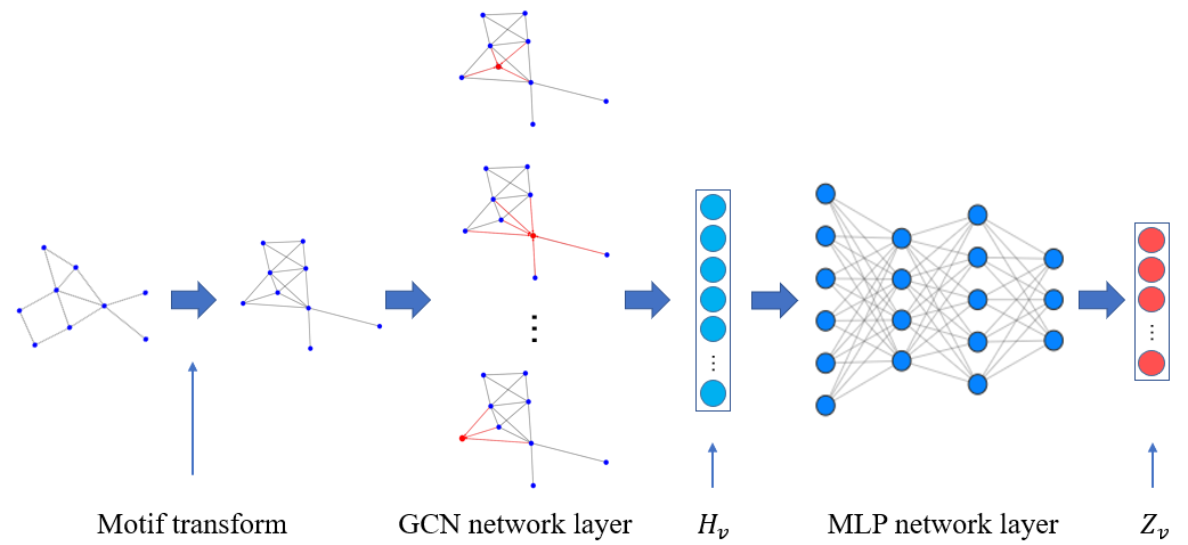}
\end{center}
\caption{mGCMN Model.~$H_v$  represents the middle embedded representation and $Z_v$  represents the final embedded representation.}
\end{figure*}

\subsubsection{Embedding Approaches Based on Random Walk}
This type of method is also a type of direct coding, where the key innovation is to optimize node embeddings. Instead of using deterministic graphical proximity measures, this kind of method uses flexible, random graphical proximity measures (essentially, it is the frequency of node pairs appearing in the same random walks), which performs well in many scenarios. The representative methods are node2vec$^{6}$, DeepWalk$^{5}$ and HARP$^{15}$, etc. Node2vec creatively uses two hyperparameters (backtracking parameter $p$ and forward parameter $q$) to control random walk, making it a compromise between depth-first random walk and breadth-first random walk. And Deepwalk is another famous approach based on random walks, which uses truncated random walk to convert the non-linear graph structure into multiple linear sequences of nodes. As for HARP, a process called graph coarsening is used in this method, which merges closely related nodes in graph $G$ into "super nodes" , and then DeepWalk, node2vec or other methods is run on the formed new graph.

\subsubsection{Embedding Approaches Based on Neural Network}
The above two types of node embedding methods are direct encoding methods. However, these direct encoding methods independently generate a representation vector for each node trained, which leads to many disadvantages: i) no shared parameters between nodes; ii) high computational complexity; iii) failing to leverage node attributes during encoding; iv) only for known nodes. This leads to the emergence of a neural network-based node representation method, which overcomes the above disadvantages and achieves excellent results in many aspects. The representative methods are Deep Neural Graph Representations$^{16}$ (DNGR), Structural Deep Network Embeddings$^{17}$ (SDNE), Graph Neural Network (GNN) and Graph Convolutional Network (GCN), etc. The DNGR and SDNE methods reduce the computational complexity, which use deep learning methods (autoencoder$^{18}$) to compress the relevant information of the node's local neighbors. And GNN is an original graph neural network which implements a function that maps the graph and one of its nodes to Euclidean space. As for GCN, it is a very well-known method first proposed by Kipf et al.$^{8}$. In this method, the convolution operation (representing any node as a function of its neighborhood, like convolutional neural network in the field of image processing) is cleverly applied to the graph structure.

\section{Method}

The key idea of our method is that we believe that combining motifs is in essence to redefine the node neighbors and redistribute the weight of the graph network. And we regard the graph convolution network combined with motif as a pre-processing tool. In general, we combine the custom motif matrix M with the graph convolution network to process the nodes' local neighborhood feature information (for example, the nodes' text attributes, statistical properties), and pass the result into the fully connected network to get the final classification result. The process is shown in Fig.2.

First, the custom motif matrix M converts the original edge adjacency graph into a motif adjacency graph (equivalent to redefining the weight); then, the graph convolution operation is performed on the motif adjacency graph to obtain the middle embedded representation of each point; finally, the intermediate embedded representation is input to a fully connected network for further processing to obtain the final embedded representation.

Next, we first introduce the custom motif matrix $M$ and various motifs in Section 3.1; then, we describe the mGCMN embedding algorithm to generate embeddings for nodes in Section 3.2; finally in Section 3.3, we give complexity analysis of the algorithm and make a proof at the same time.

\subsection{Motif Matrix}

\begin{figure}[h]
\center
\includegraphics[width=\linewidth]{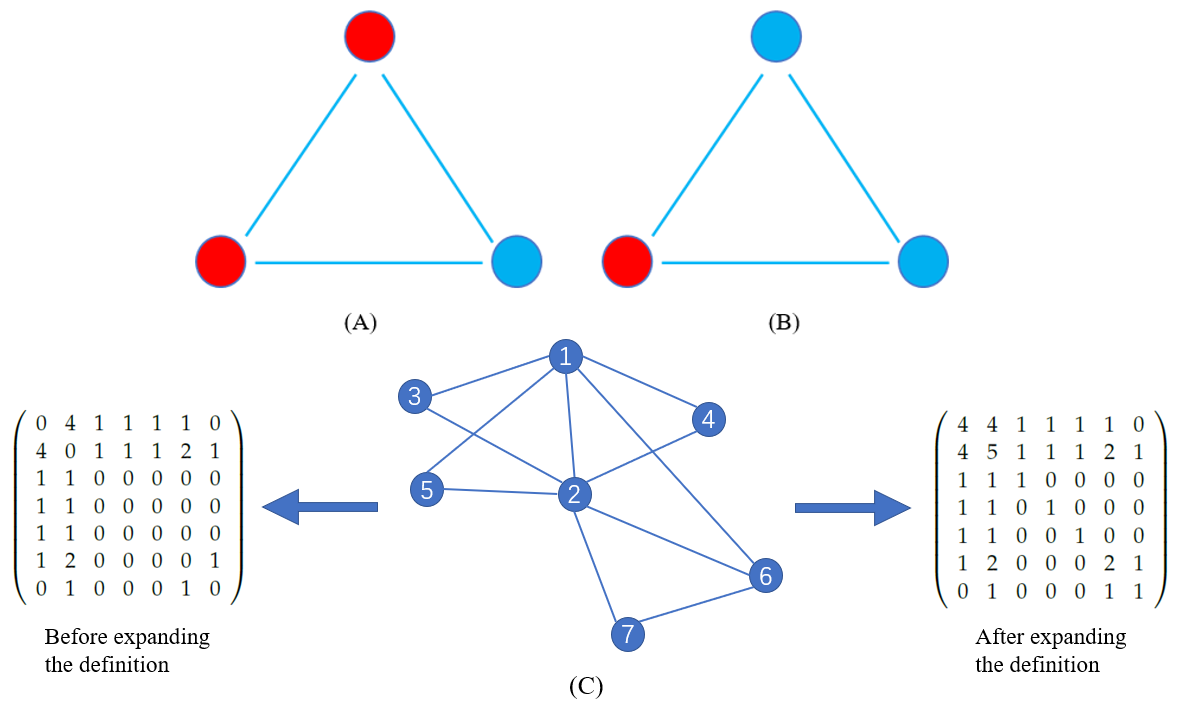}
\caption{Example.~Relative position relationships are marked by red nodes in (A) and (B).}
\end{figure}

We will explain motif matrix in detail in this section. For the convenience of explanation, some commonly used symbols are agreed on. Formally, let graph $G=(V,E)$, where $V$ is the set of the nodes in network, and $E$ is the set of the edges, $E \subseteq (V \times V)$. Given a labeled network with node feature information $G=(V,E,X,Y)$, where $X \in R^{(N \times T)}$ ($N$ is the number of the nodes in network, $T$ is the feature dimension) is the feature information matrix and $Y \in R^{(N \times L)}$ ($L$ is the feature dimension) is the label information matrix, our goal is to use the labels of some of the nodes for training, and generate a vector representation matrix $Z$ of the nodes.

Then, we give:

\noindent\textbf{Definition 1:} Given a graph $G=(V,E,X,Y)$, a motif $M$ with central node $v_M$ (the node currently being followed) is defined as $M=(V_M,E_M,v_M)$, where $V_M$ is the node set of $M$ containing $v_M$, and $E_M \subseteq E$ satisfies that $\forall v,u \in V_M$, if $(v,u)$ is in $M$, then $(v,u) \in E_M$.

\noindent\textbf{Definition 2:} An instance $S_u=(V_S,E_S)$ of motif $M$ with central node $u$ on graph $G=(V,E,X,Y)$ is a subgraph of $G$, where $V_S \subseteq V$ and $E_S \subseteq E$, satisfying (i)  $u \in V_S$, and $\Phi(u)=v_M$; (ii) $\forall a,b \in V_S$, if $(\Phi(a),\Phi(b))\in E_M$, then $(a,b)\in E_S$, where $\Phi: V_S \rightarrow V_M$ is an arbitrary bijection.

After that, we can define the motif matrix. Given a motif $M$, a motif matrix $A^{M}$ of $M$ is defined as: $a_{vu}^{M}$ is the number on the $v$-row and $u$-column of the motif matrix $A^{M}$, and it is equal to the number of the times that nodes $v$ and $u$ appear in the same instances of $M$. Formally,$$a_{vu}^{M}=\sum_{E_S}I((v,u) \in E_S),$$

\noindent where $I(\cdot)$ is indicative function.

The above is the usual definition of motif adjacency matrix, and particularly, we extend it. When the nodes $v$ and $u$ come to be the same node, we also record its number of the times that appearing in the same instances of $M$. For example, the positional relationship of the motif "triangle" is like Fig.3 (A), while after expanding the definition, the position relationship like Fig.3 (B) is also counted. For a specific example, its triangle motif adjacency matrix is shown in Fig.3 (C).

\subsection{Embedding Algorithm}

For specific algorithm, see: mGCMN embedding algorithm.

\begin{algorithm}[htb]  
	\caption{\textbf{:} mGCMN embedding algorithm.}  
	\label{alg:Framwork}  
	\begin{algorithmic}[1]  
    	\Require  
      		Motif adjacency matrix $A^{M}$ ($M$ is the corresponding motif);
			Node feature matrix $X$;
			The number of motif-based GCN layers $H_1$;
			The number of Multi-Layer Perceptron (MLP) layers $H_2$; 
    	\Ensure  
     	 	Representation matrix $Z$;  
    	\State $h_v^{0}=x_v, \forall v \in V$; 
		\For{$k=1...H_1$}
			\For{$v \in V$}
				\State $h_v^{k} \leftarrow f(h_v^{k-1},h_u^{k-1}),u \in N_M(v)$;
			\EndFor
		\EndFor
		\For{$k=1...H_2$}
			\For{$v \in V$}
				\State $h_v^{k+H_1} \leftarrow \tilde{f}(h_v^{k+H_1-1})$;
			\EndFor
		\EndFor
    	\State $z_v=h_v^{(H_1+H_2)}, \forall v \in V$;
    	\State \textbf{return} $Z$ (whose column vectors are $z_v, v \in V$); 
	\end{algorithmic}  
\end{algorithm}

We first obtain a custom motif matrix as defined in Section 3.1. The numbers of GCN layers based on motif and MLP layers are specified by the users in advance; The initialization of all nodes is expressed as : $h_v^{0}=x_v, \forall v \in V$, in line 1; In lines 2-6, we perform a graph convolution operation based on motif, in the formula $h_v^{k} \leftarrow f(h_v^{k-1},h_u^{k-1}),u \in N_M(v)$, $f( \cdot )$ represents a weighted nonlinear aggregation function, whose purpose is to reorganize the information of the target node and its neighbors. Formally,

\begin{center}
$h_v^{k}=\sigma(\sum_{u \in N_M(v) \cup \{v\}}a_{vu}^{M} \cdot h_u^{k-1} \cdot W_k),$
\end{center}

\noindent where $h_v^{k}$ is the hidden representation of node $v$ in the $k$-th layer; $a_{vu}^{M}$ is the number on the $v$-row and $u$-column of the motif matrix $A^{M}$, indicating the closeness between nodes of $v$ and $u$; $W_k$ is the parameter matrix to be trained of layer $k$; $N_M(v)$ is the neighborhood nodes set of node $v$ in the motif matrix $A^{M}$; $\sigma( \cdot )$ represents for ReLU function.

In lines 7-11, the processing results of the graph convolution operation based on motif are sent to MLP for further processing, in the formula $h_v^{k+H_1} \leftarrow \tilde{f}(h_v^{k+H_1-1})$, $\tilde{f}( \cdot )$ represents a non-linear activation unit that further processes the information of target nodes. Formally, we have

\begin{center}
$h_v^{k+H_1}=\tilde{\sigma}(h_v^{k+H_1-1} \cdot W_{k+H_1}),$
\end{center}

\noindent where $\tilde{\sigma}( \cdot )$ represents for ReLU function, except for the last layer (in the last layer, $\tilde{\sigma}( \cdot )$ represents for Softmax function).

Then, the final representation vector $z_v$ of node $v$ is obtained. Finally, the cross entropy function is used as the loss function to train the parameters of our model:

\begin{center}
$loss = \sum_v(y_v \cdot log (z_v) + (1-y_v \cdot log (1-z_v)), v \in trainset,$
\end{center}
where $y_v$ is the label of the node $v$.

\subsection{Complexity Analysis}

Our method is based on GCN. And from the related work of Kipf et al.$^{8}$, we know that the computational complexity of the original GCN based on the following formula is $\mathcal{O}(|\mathcal{E}|CHF)$, where $\mathcal{E}$ is the edge set of the graph:
\begin{center}
$Z = f(X,A) =$ softmax$\left( \hat{A} ~ {\rm ReLU}\left(\hat{A}XW^{(0)}\right) W^{(1)}\right),$
\end{center}
Here, $A$ is the adjacency matrix and $X$ is the feature matrix. And $\hat{A}$ is the normalized processing matrix of the adjacency matrix $A$. $W^{(0)} \in \mathbb{R}^{C \times H}$ is an input-to-hidden weight matrix and $W^{(1)} \in \mathbb{R}^{H \times F}$ is a hidden-to-output weight matrix, where $C$ is input channels, $H$ is feature maps in the hidden layer and $F$ is feature maps in the output layer.$^{8}$
Next, we will prove that the calculation complexity of our method is also $\mathcal{O}(|\mathcal{E}|CHF)$ while keeping the number of hidden layers unchanged and using the Motif matrix instead of the original adjacency matrix.
\begin{proof}
Let $D$ be the maximum degree of the nodes in graph $G$, $N$ denote the number of nodes in $G$, $M$ denote the Motif matrix and $A$ denote the original adjacency matrix.

For the triangle Motif, consider the zero element in A. Let us assume that $A_{i,j}$ is $0$, that is, there is no edge between nodes $i$ and $j$. So we can know $M_{i,j} = 0$ (nodes $i$,$j$ are not in the same triangle). So for the triangle Motif, the computational complexity does not change;

For the wedge Motif, we consider $A^2$. If $A_{i,j}^2 = 0$, it means that node $i$ can not reach node $j$ by 2 steps (i.e., node $i$ is not a second-order neighbor of node $j$), which means that the nodes i and j are not in the same wedge. So we can know $M_{i,j} = 0$. Then consider the number of non-zero elements in the matrix $A^2$, which is set to $n$. According to $A_{i,}^2 = A_{i,} \cdot A$ ($A_{i,}$ represents the $i$-th row of $A$), we can know that $n_{i,} \le d_i \cdot D$, where $n_{i,}$ represents the number of non-zero elements in $A_{i,}^2$ and $d_i$ is the degree of node $i$. Therefore, the total number $n$ of non-zero elements in $A^2$ satisfies equation:
\begin{center}
$~n \le d_1 \cdot D + d_2 \cdot D + ... + d_N \cdot D$\\
\end{center}
\begin{center}
$~= (d_1 + d_2 + ... + d_N) \cdot D~~~~~$
\end{center}
\begin{center}
$= 2 |\mathcal{E}| D.~~~~~~~~~~~~~~~~~~~~~~~~~~~~~~~$
\end{center}

So the number of non-zero elements in $M$ is no more than $2 |\mathcal{E}| D$. Then the computational complexity is $\mathcal{O}(2 |\mathcal{E}| DCHF) = \mathcal{O}(|\mathcal{E}|CHF)$.
\end{proof}

\section{Experiments}
In section 4.1, we introduce the datasets used in the experiment, and the specific settings of the experiment are described in section 4.2.

\subsection{Datasets}

\begin{table}[h] 
	\caption{Datasets Statistics.}
	\centering
	\begin{tabular}{l l c c c c}
		\toprule
		Datasets & Types & Nodes & Edges & Features & Classes \\
		\midrule
		Cora & Citation & 2708 & 5429 & 1433 & 7 \\
		Citeseer & Citation & 3327 & 4732 & 3703 & 6 \\
		Pubmed & Citation & 19717 & 44338 & 500 & 3 \\
		107Ego & Social & 1045 & 53498 & 576 & 9 \\
		414Ego & Social & 159 & 3386 & 105 & 7 \\
		1684Ego & Social & 792 & 28048 & 319 & 16 \\
		1912Ego & Social & 755 & 60050 & 480 & 46 \\
		\bottomrule
	\end{tabular}
\end{table}

The statistics of the experimental datasets are shown in the Table 1. In the citation network dataset (Citeseer, Cora, and Pubmed), nodes represent documents and edges represent citation links; In the social network dataset (Ego-Facebook), nodes represent users and edges represent interactions between users.

\noindent\textbf{Citation Network Datasets: Citeseer, Cora, and Pubmed.} The three citation network datasets contain a sparse feature vector for each document and a list of reference links between the documents. Citation links are considered as (undirected) edges and each document has a category label$^{8}$.

\noindent\textbf{Social Network Dataset: Ego-Facebook.} This dataset consists of 'circles' (or 'friends lists') from Facebook (Facebook data was collected from survey participants). There are many subsets of the Ego-Facebook dataset. Take '107Ego' as an example. The dataset includes node features, edge sets, node category sets, and self-networks (network with node 107 as the core), where each user is considered as a node, the interaction is considered as an (undirected) edge, and each user has a feature attribute vector and a category label. We choose the suitable Ego-Facebook subsets for experiments, and after preprocessing, the data whose information has been lost is removed.

\subsection{Experimental setup}
We first add motif of the general sense to the GCN network, the purpose is to observe whether the individual motif will work, and if it works, which kind of motif works better; then the general sense motif is changed to a custom motif, and connected to the MLP network for further processing (that is, the complete mGCMN algorithm), and the method is also marked as mGCMN.

When we perform experiments on the citation network dataset (Citeseer, Cora, and Pubmed), the number of GCN network layer and MLP network layers are set to 1 and 0 respectively for Citeseer and Pubmed; as for Cora, the number of GCN network layer and MLP network layers are set to 2 and 1 respectively; and the other parameters are set as the settings in [8]. When performing experiments on the social network dataset (Ego-Facebook), the number of GCN network layer is set to 2, and the number of MLP network layers is set to 1.

In all of the experiment, many motifs have been experimented, and finally a mixed matrix of motifs and graph adjacency matrix are choosen, which will get the best results. The mixed matrix parameter $\lambda$ is determined by grid search. The details are as follows.

In the mGCMN method: the ratio of edge, triangle motif and wedge motif is 8: 1: 2 on the Citeseer dataset; the ratio of edge, triangle motif and wedge motif is 8: 1: 3 on the Cora dataset; the ratio of edge and wedge motif is 9: 1 on the Pubmed dataset; the ratio of edge and wedge motif is 9: 1 on the Facebook-107Ego dataset; the ratio of edge and triangle motif is 4: 1 on the Facebook-414Ego dataset; the ratio of edge and wedge motif is 1: 1 on the Facebook-1684Ego dataset; the ratio of edge and wedge motif is 4: 1 on the Facebook-1912Ego dataset.

Finally, the prediction vector is used to compare the prediction accuracy of classification on the test set. See specific procedures in our program.

\section{Results and Discussion}
In this section, we introduce a variety of baseline methods, and show the comparison of all experimental results as follows:

\subsection{Experimental Results on Citation Network Datasets (Citeseer, Cora, Pubmed)}
First, we use the citation network datasets (Citeseer, Cora, and Pubmed) for the experiment, and compare the experimental results with various baseline methods$^{8}$. The results are shown in Table 2.

\begin{table}[h] 
	\caption{The results of classification accuracy for various baseline methods and mGCMN.}
	\centering
	\begin{tabular}{c c c c}
		\toprule
		\textbf{Method} & \textbf{Cora} & \textbf{Citeseer} & \textbf{Pubmed} \\
		\midrule
		ManiReg & 59.5 & 60.1 & 70.7 \\
		SemiEmb & 59.0 & 60.1 & 71.1 \\
		LP & 68.0 & 45.3 & 63.0 \\
		DeepWalk & 67.2 & 43.2 & 65.3 \\
		ICA & 75.1& 69.1 & 73.9 \\
		Planetoid & 75.7 & 64.7 & 77.2 \\
		GCN & 81.5 & 70.3 & 79.0 \\
		\midrule
		mGCMN & \textbf{82.3} & \textbf{71.8} & \textbf{79.5} \\
		\midrule
		\textbf{CC} & 0.09350 & 0.14297 & 0.05380 \\
		\bottomrule
	\end{tabular}
\end{table}

The table shows the comparative results of our method with the methods of label propagation (LP)$^{19}$, semi-supervised embedding (SemiEmb)$^{20}$, manifold regularization (ManiReg)$^{21}$, iterative classification algorithm (ICA)$^{22}$ and Planetoid$^{23}$, and DeepWalk is a method based on random walks, as stated at the beginning of the article, whose sampling strategy can be seen as a special case of node2vec with $p=1$ and $q=1$. As for method named GCN, which is the first method to achieve convolution on the graph, it is the best performing baseline method, and we can see that our method performs better than it on every dataset.

One interesting finding is that we noticed that the global clustering coefficients (CC) of these three graph networks are 0.09350, 0.14297 and 0.05380, whose ordering is consistent with the order of the increase of our method (compared to GCN). A more intuitive display is shown in the Fig.4 (a). In the next experiment, this phenomenon appears again, and we believe that this illustrates the rationality of our application of higher-order neighborhood information. The clustering coefficient $CC$ is calculated according to the following formula:

\begin{center}
$$CC = \frac{3 * number ~ of ~ triangles}{number ~ of ~ wedge}.$$
\end{center}

\begin{figure}
\center
\subfigure[]{
\includegraphics[width=\linewidth]{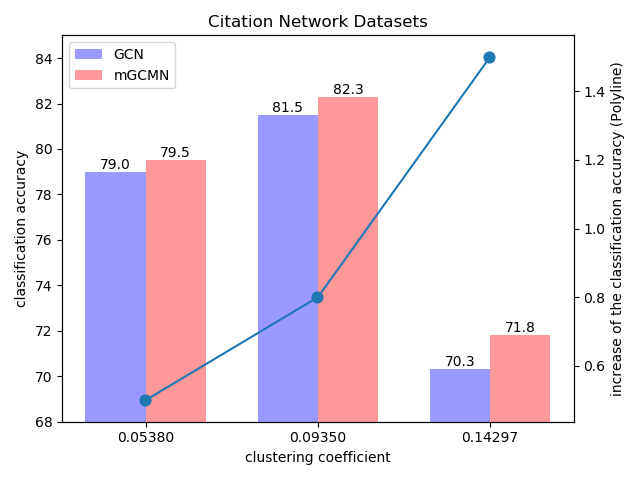}}

\subfigure[]{
\includegraphics[width=\linewidth]{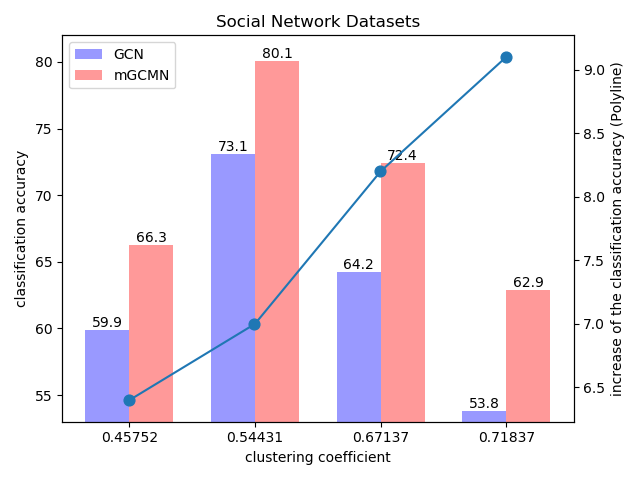}}
\caption{Relationship between (increase of) the classification accuracy and the clustering coefficient.}
\end{figure}

\subsection{Experimental Results on Social Network Dataset (Ego-Facebook)}
Now, we use the social network dataset (Ego-Facebook) for the experiments, and compare the experimental results with the classic method based on random walks $--$ Deepwalk$^{5}$, and the best performing baseline method $--$ GCN$^{8}$. The experimental results are shown in the Table 3:

\begin{table}[h] 
	\caption{The results of classification accuracy for Deepwalk, GCN, and mGCMN.}
	\centering
	\begin{tabular}{c r r r r}
		\toprule
		\textbf{Method} & \textbf{107Ego} & \textbf{414Ego} & \textbf{1684Ego} & \textbf{1912Ego} \\
		\midrule
		DeepWalk & (77.5) & (79.2) & (64.4) & (66.5) \\
		GraRep & (90.0) & (85.4) & (76.3) & (77.0) \\
		GCN & 73.1(92.5) & 64.2(93.8) & 59.9(81.9) & 53.8(77.0)\\
		\midrule
		mGCMN & \textbf{80.1(95.0)} & \textbf{72.4\ (100)} & \textbf{66.3(88.8)} & \textbf{62.9(84.0)} \\
		\midrule
		\textbf{CC} & 0.54431 & 0.67137 & 0.45752 & 0.71837 \\
		\bottomrule
	\end{tabular}
\end{table}

In the Table 3, GraRep$^{13}$ works by defining a more accurate loss function that allows non-linear combinations of different local relationship information to be integrated.

Among them, GCN and mGCMN require random weight initialization, so the average accuracy of 100 runs after the random weight initialization is reported, the highest accuracy in all experiments is shown in brackets. Deepwalk and GraRep does not require random weight initialization, so the best performance was reported after the hyperparameters were determined.

As can be seen from Table 3, the experimental results of both parts of our method are significantly higher than the other baseline. And we again see that the ranking of the global clustering coefficients is consistent with the ranking of the improvement of our method (compared to GCN). A more intuitive display is shown in the Fig.4 (b). We think this phenomenon may help to search the data which is suitable to process using our method.

\section{Conclusion}
In this paper, we have designed a new framework combined with motifs $--$ mGCMN, which can effectively aggregate node information (we think it can be seen as accomplishing this by defining a new neighborhood structure), and capture higher-order features through deeper learning. The results have shown that mGCMN can effectively generate embeddings for nodes of unknown category and is always better than the baseline methods. At the same time, the experiment also reveal the relationship between increase of the classification accuracy and the clustering coefficient.

There are many extensions and potential improvements to our method, such as further exploring the relationship between motifs and graph statistics and extending mGCMN to handle directed or multi-graph mode. Another interesting direction for future work is to explore how to use the adjacency matrix more efficiently and flexibly.


%



\ifCLASSOPTIONcompsoc
  \section*{Acknowledgments}
\else
  \section*{Acknowledgment}
\fi

This work is supported by the Research and Development Program of China（No.2018AAA0101100）, the Fundamental Research Funds for the Central Universities,   the International Cooperation Project No.2010DFR00700, Fundamental Research of Civil Aircraft No. MJ-F-2012-04 and the Beijing Natural Science Foundation (1192012, Z180005).

\ifCLASSOPTIONcaptionsoff
  \newpage
\fi


\begin{thebibliography}{1}

\bibitem{IEEEhowto:kopka}
L. Backstrom and J. Leskovec, "Supervised random walks: predicting and recommending links in social networks,"
\emph{Proceedings of the Fourth ACM International Conference on Web Search and Data Mining} (WSDM), 2011.
\vspace{1.5ex}
\bibitem{IEEEhowto:kopka}
M. Zitnik and J. Leskovec, "Predicting multicellular function through multi-layer tissue networks," \emph{Bioinformatics}, Vol. 33, no. 14, 2017, pp. l190-l198.
\vspace{1.5ex}
\bibitem{IEEEhowto:kopka}
X. Wang, Y. Ye and A. Gupta, "Zero-shot Recognition via Semantic Embeddings and Knowledge Graphs," \emph{Computer Science}, 2018.
\vspace{1.5ex}
\bibitem{IEEEhowto:kopka}
M. Belkin and P. Niyogi, "Laplacian eigenmaps and spectral techniques for embedding and clustering," \emph{NIPS}, 2002.
\vspace{1.5ex}
\bibitem{IEEEhowto:kopka}
B. Perozzi, R. Al-Rfou, and S. Skiena, "Deepwalk: Online learning of social representations," \emph{Computer Science}, 2014.
\vspace{1.5ex}
\bibitem{IEEEhowto:kopka}
A. Grover and J. Leskovec, "node2vec: Scalable feature learning for networks," \emph{Computer Science}, 2016.
\vspace{1.5ex}
\bibitem{IEEEhowto:kopka}
F. Scarselli et al., "The Graph Neural Network Model," \emph{IEEE Transactions on Neural Networks}, vol.20, no.1, 2009, pp. 61-80.
\vspace{1.5ex}
\bibitem{IEEEhowto:kopka}
T.N. Kipf and M. Welling, "Semi-supervised classification with graph convolutional networks," \emph{ICLR}, 2016.
\vspace{1.5ex}
\bibitem{IEEEhowto:kopka}
R. Milo et al., "Network Motifs: Simple Building Blocks of Complex Networks," \emph{Science}, vol.298, 2002, pp. 824-827.
\vspace{1.5ex}
\bibitem{IEEEhowto:kopka}
A.R. Benson, D.F. Gleich, and J. Leskovec, "Higher-order organization of complex networks," \emph{Science}, 2016, pp. 163-166.
\vspace{1.5ex}
\bibitem{IEEEhowto:kopka}
W.L. Hamilton, R. Ying, and J. Leskovec, "Representation Learning on Graphs: Methods and Applications," \emph{Computer Science}, 2017.
\vspace{1.5ex}
\bibitem{IEEEhowto:kopka}
A. Ahmed et al., "Distributed large-scale natural graph factorization," \emph{IW3C2 - International World Wide Web Conference}, 2013.
\vspace{1.5ex}
\bibitem{IEEEhowto:kopka}
S. Cao, W. Lu, and Q. Xu, "Grarep: Learning graph representations with global structural information," \emph{ACM}, 2015, pp. 891-900.
\vspace{1.5ex}
\bibitem{IEEEhowto:kopka}
M. Ou et al., "Asymmetric transitivity preserving graph embedding," \emph{22nd ACM SIGKDD Conference on Knowledge Discovery and Data Mining} (KDD), 2016.
\vspace{1.5ex}
\bibitem{IEEEhowto:kopka}
H. Chen et al., "Harp: Hierarchical representation learning for networks," \emph{Computer Science}, 2017.
\vspace{1.5ex}
\bibitem{IEEEhowto:kopka}
S. Cao, W. Lu, and Q. Xu, "Deep neural networks for learning graph representations," \emph{AAAI}, 2016.
\vspace{1.5ex}
\bibitem{IEEEhowto:kopka}
D. Wang, P. Cui, and W. Zhu, "Structural deep network embedding," \emph{KDD}, 2016.
\vspace{1.5ex}
\bibitem{IEEEhowto:kopka}
G.E. Hinton and R.R. Salakhutdinov, "Reducing the dimensionality of data with neural networks," \emph{Science}, vol.313, 2006, pp. 504-507. 
\vspace{1.5ex}
\bibitem{IEEEhowto:kopka}
X. Zhu, Z. Ghahramani, and J.D. Lafferty, "Semi-supervised learning using gaussian fields and harmonic functions," \emph{International Conference on Machine Learning}, vol.3, 2003, pp. 912-919.
\vspace{1.5ex}
\bibitem{IEEEhowto:kopka}
J. Weston et al., "Deep learning via semi-supervised embedding," \emph{Neural Networks: Tricks of the Trade}, 2016, pp. 639-655.
\vspace{1.5ex}
\bibitem{IEEEhowto:kopka}
M. Belkin, P. Niyogi, and V. Sindhwani, "Manifold regularization: A geometric framework for learning from labeled and unlabeled examples," \emph{Journal of machine learning research}, vol.7, 2006, pp. 2399-2434.
\vspace{1.5ex}
\bibitem{IEEEhowto:kopka}
Q. Lu and L. Getoor, "Link-based classification," \emph{International Conference on Machine Learning}, vol.3, 2003, pp. 496-503.
\vspace{1.5ex}
\bibitem{IEEEhowto:kopka}
Z. Yang, W.W. Cohen, and R. Salakhutdinov, "Revisiting semi-supervised learning with graph embeddings," \emph{International Conference on Machine Learning}, 2016.
\vspace{1.5ex}

\end{thebibliography}
\end{document}